\documentclass[11pt,a4paper]{article}

\usepackage{geometry}
\usepackage{authblk}

\usepackage[scheme=plain]{ctex} 
\usepackage{amsmath}
\usepackage{graphicx}
\usepackage{ragged2e}
\usepackage{subcaption}
\usepackage{amssymb}
\usepackage{bm}
\usepackage{hyperref}

\hypersetup{
    colorlinks=true,
    linkcolor=blue,
    citecolor=blue,
    urlcolor=blue
}

\begin{document}
\justifying % 全局双端对齐

\title{Study of $\Lambda$ \textit{CP} Asymmetry Parameter and Electric Dipole Moment at the Super Tau-Charm Facility}

% 1. 定义作者（中括号内填单位序号，大括号内填姓名与角标说明）
\author[1]{Mingyu Yu (于明玉)}
\author[2,3]{Jianyu Zhang (张剑宇)$^{\dagger}$}
\author[2]{Jinlin Fu (傅金林)}
\author[1]{Xingtao Huang (黄性涛)$^{\ddagger}$}

% 2. 定义单位（中括号内填对应的单位序号）
\affil[1]{Institute of Frontier and Interdisciplinary Science, Key Laboratory of Particle Physics and Particle Irradiation, Shandong University, Qingdao, Shandong 266237, People's Republic of China}
\affil[2]{School of Physical Sciences, University of Chinese Academy of Sciences, Beijing 100049, People's Republic of China}
\affil[3]{National Centre for Nuclear Research, Pasteura 7, 02-093 Warsaw, Poland}

% 3. 定义日期（留空则不显示生成当天的日期）
\date{}

% 4. 【关键】全部信息定义完毕后，再执行 \maketitle 渲染！
\maketitle

\let\thefootnote\relax\footnotetext{* This work is supported by the National Natural Science Foundation of China (NSFC) under Contracts No. 12341504, No. 12341501, No. 12341503; the National Key R\&D Program of China under Contracts No. 2022YFA1602200 and No. 2023YFA1607200; the international partnership program of the Chinese Academy of Sciences Grant No. 211134KYSB20200057; the Polish National Science Centre (Grant No. 2024/53/B/ST2/00975).}
\let\thefootnote\relax\footnotetext{$^\dagger$ E-mail: zhangjianyu@ihep.ac.cn}
\let\thefootnote\relax\footnotetext{$^\ddagger$ E-mail: huangxt@sdu.edu.cn}

\begin{abstract}
Searching for new mechanisms of \textit{CP} violation beyond the Standard Model is essential to explain the observed matter-antimatter asymmetry of the universe. By analyzing the full angular distribution of the quantum-entangled $\Lambda\overline{\Lambda}$ pairs from $J/\psi$ decays, the preliminary sensitivity of \textit{CP} asymmetry parameter $A_{CP}$ and Electric Dipole Moment of $\Lambda$ is studied at the future Super Tau-Charm Facility (STCF). Based on $3.4 \times 10^{12}$ $J/\psi$ events at STCF, the projected sensitivity for $A_{CP}^{\Lambda}$ reaches the $10^{-4}$ level, while the $\Lambda$ EDM can be probed at $10^{-20}$~$e\text{ cm}$. These results highly indicate that STCF will provide an unprecedented opportunity to search for new mechanisms of \textit{CP} violation with an expected world-leading precision.
\end{abstract}

\vspace{0.8em}

\section{INTRODUCTION}
The significant asymmetry between matter and antimatter is observed in our universe~\cite{Fields:2019pfx}. According to Sakharov's criteria, a sufficiently large source of \textit{CP} violation is indispensable to drive this matter dominant universe~\cite{Sakharov:1967dj}. As the bedrock theory of particle physics, the Standard Model (SM) provides a comprehensive description of elementary particles and their interactions. To date, the \textit{CP} violation observed experimentally in the $K$~\cite{PhysRevLett.13.138}, $B$~\cite{Aubert:2001nu, Abe:2001xe}, $D$~\cite{LHCb:2019hro} meson systems and the $\Lambda^0_b$ baryon system~\cite{LHCb:2025baryonCPV} agree well with the SM predictions. However the magnitude of \textit{CP} violation predicted by SM is too small to explain the matter-antimatter asymmetry. Therefore, searching for new mechanisms of \textit{CP} violation is of great importance.

The permanent electric dipole moment (EDM) of particles provides a pristine and
unambiguous probe for this~\cite{Pospelov:2005pr}. The existence of non-zero EDM directly implies the simultaneous violation of both parity (\textit{P}) and time reversal (\textit{T}) symmetries, which also means \textit{CP} violation under the assumption of \textit{CPT} conservation~\cite{Pospelov:2005pr}. Furthermore, the SM predictions for hadronic EDMs arising from the CKM phase and the QCD vacuum angle $\bar{\theta}$ are very small. Thus, the EDM will serve as a highly sensitive probe for investigating \textit{CP} violation and exploring new physics~\cite{Beacham:2019nyx}.  

Up to now, many extensions of the SM, such as supersymmetry (SUSY)~\cite{Fischler:1992ha} and left-right symmetric models~\cite{Frere:1991qm}, naturally predict non-zero EDMs that could be detected in the new round of experiments. Various advanced experimental techniques based on spin precession have been deployed to perform high-precision EDM measurements across nucleonic, atomic, and leptonic systems~\cite{Chupp:2017rkp}. For hyperons, Fermilab established the upper limit for the $\Lambda$ EDM at $1.5 \times 10^{-16}$~$e\text{ cm}$~\cite{Pondrom:1981mj}, which is relatively large compared to the aforementioned‌ systems. However, limited by the short lifetimes of hyperons and the technical bottlenecks inherent to fixed-target experiments~\cite{Pondrom:1981mj}, further sensitivity improvements remain difficult. Since various theoretical models yield distinct contributions to particle EDMs, high-precision EDM measurements across diverse systems serve as a powerful tool to constrain the parameter space. Moreover, if new physics couples preferentially to the strange ($s$) quark, hyperons will obtain significant EDM effects, offering an unique opportunity to probe the underlying flavor structure of new physics. Thus, further experimental exploration of hyperon EDMs is crucial~\cite{Pospelov:2005pr, Chupp:2017rkp}.

Beyond static properties like the EDM, testing the direct \textit{CP} violation in hyperons non-leptonic weak decays is equally essential. In 1986, Donoghue and Pakvasa proposed that the observable quantity of \textit{CP} violation could be constructed using the parity-violating parameters of the hyperon ($\alpha$) and antihyperon ($\bar{\alpha}$)~\cite{Donoghue:1985pv, Donoghue:1986yc}:
\begin{equation}
\label{eq:acp_alpha}
A_{CP} = \frac{\alpha + \bar{\alpha}}{\alpha - \bar{\alpha}},
\end{equation}
where $\alpha = \frac{2\text{Re}(s^*p)}{|s|^2 + |p|^2}$ when studying the decays of a spin-1/2 hyperon into a final state comprising a spin-1/2 baryon and a spin-0 pseudoscalar meson, and $s$ and $p$ denote the parity-violating and parity-conserving wave amplitudes of the weak decay~\cite{Lee:1956qn}. As a $CP$-odd observable, a non-zero $A_{CP}$ directly signals the \textit{CP} violation~\cite{Okubo:1958zz, Pais:1959zz}. Since the SM predicts $A_{CP} \sim \mathcal{O}(10^{-5}$--$10^{-4})$~\cite{Donoghue:1985pv, Donoghue:1986yc}, it is also sensitive to new physics.

To efficiently study both $A_{CP}$ and the hyperon EDMs, a comprehensive full angular analysis method utilizing the subtle angular distributions differences of the entangled hyperon-antihyperon pairs has been proposed. And the parameters of interest can be simultaneously extracted by fitting the five-dimensional joint angular distribution of the hyperon-antihyperon pair and their subsequent decay products~\cite{Fu:2023, He:1993, He:1994}. An electron-positron ($e^+e^-$) collider provides an ideal platform for such studies, where the substantial $J/\psi$ production cross section and its large branching fraction to hyperon-antihyperon pairs yield abundant, clean data samples~\cite{BESIII:2020nme, Li:2016tby, STCF:2024}. More importantly, pair-produced hyperons from charmonium decays are inherently spin-entangled, which significantly enhances the experimental sensitivity for detecting \textit{CP} violating signals. Based on the full angular analysis method, the BESIII experiment achieved a major milestone by constraining the $\Lambda$ EDM to the $\mathcal{O}(10^{-19})$~$e\text{ cm}$ level~\cite{BESIII:2025edm}, representing a three-order-of-magnitude improvement over the previous fixed-target results. Simultaneously, the precision of $A_{CP}^{\Lambda}$ was determined at the $\mathcal{O}(10^{-3})$ level. However, no $\Lambda$ EDM and \textit{CP} violation have been observed. Searching for tiny \textit{CP} violation signals beyond the SM still requires larger statistics. 

Designed as a high-luminosity symmetric $e^+e^-$ collider, the future Super Tau-Charm Facility (STCF) is expected to provide $3.4 \times 10^{12}$ $J/\psi$ events annually~\cite{STCF:2024}. With this larger statistics, much more precise measurements of hyperons \textit{CP} violating can be achieved. In this paper, we evaluate the projected sensitivities of the $\Lambda$ EDM and $A_{\mathit{CP}}^{\Lambda}$ for the decay chain $J/\psi \rightarrow \Lambda(\rightarrow p\pi^-)\bar{\Lambda}(\rightarrow \bar{p}\pi^+)$ based on STCF Monte Carlo (MC) simulation. In addition, preliminary studies of potential background contributions and systematic uncertainties are also performed.

\section{STCF DETECTOR AND MONTE CARLO SIMULATION}
\label{sec:stcfandmc}
The STCF is a proposed next-generation $e^+e^-$ collider. It covers a broad center-of-mass energy range of $\sqrt{s} = 2.0$--$7.0$~GeV, with a peak luminosity of $0.5 \times 10^{35}$~cm$^{-2}$s$^{-1}$ at $\sqrt{s} = 4$~GeV. Furthermore, the design reserves sufficient potential for a future luminosity upgrade and incorporates a second-phase plan to achieve beam polarization. However, beam polarization is not assumed in the present simulation study. Consequently, the STCF will serve as a golden platform to perform precision tests of the SM and explore new physics beyond it.

From the inside out, the STCF detector is preliminarily designed to consist of a tracking system which will provide precise measurements of charged particle momentum and position, a particle identification system (PID), an electromagnetic calorimeter (ECAL), a 1 T superconducting solenoid, and a muon detector (MUD). More details for each subdetector design can be found in Ref.~\cite{STCF:2024}. For the tracking system, it is composed of an inner tracker (ITK) and a main drift chamber (MDC)， where the ITK is specifically designed to optimize tracks with low-momentum and improve vertex reconstruction precision. Both cover a polar angle range of $20^\circ$ to $160^\circ$. According to physics requirements, the tracking system needs to achieve track reconstruction efficiencies exceeding 99\% for charged particles in the high transverse momentum region ($p_{\text{T}} > 300~\text{MeV}/c$) while maintaining 90\% efficiency at $p_{\text{T}} = 100~\text{MeV}/c$. With this unprecedented set of performance specifications, the STCF will provide a large solid-angle coverage and high tracking efficiency, guaranteeing the high-performance reconstruction of $\Lambda$ hyperons.  

This analysis is performed based on the MC sample produced within the STCF offline software framework (\textsc{OSCAR})~\cite{Huang:2023jinst}, which includes event generation, a full \textsc{Geant4}-based detector simulation, digitization, background mixing, and reconstruction. To evaluate the sensitivity of $A_{CP}^{\Lambda}$ and $\Lambda$ EDM, a signal MC sample of $3 \times 10^6$ $J/\psi \rightarrow \Lambda(\rightarrow p\pi^-)\overline{\Lambda}(\rightarrow \overline{p}\pi^+)$ events is generated according to the amplitude explicitly described in Sec.~\ref{sec:formalism}. Additionally, a uniformly distributed phase-space (PHSP) MC sample of $3 \times 10^7$ events is generated to calculate the normalization factor for the maximum likelihood fit described in Sec.~\ref{sec:fitresult}. To evaluate the potential background impact to the final analysis results, the following background samples are generated in PHSP: $1.1 \times 10^6$ events of $J/\psi \rightarrow \Lambda\overline{\Lambda}\gamma$, $3.0 \times 10^7$ events of $J/\psi \rightarrow p\pi^-\bar{p}\pi^+$, and $1.2 \times 10^5$ events for each of the $J/\psi \to \Lambda \bar{\Sigma}^0 + \mathrm{c.c.}$ channels. 

\section{FORMALISM}
\label{sec:formalism}
\begin{figure}[h!]
\centering
\includegraphics[width=0.9\textwidth]{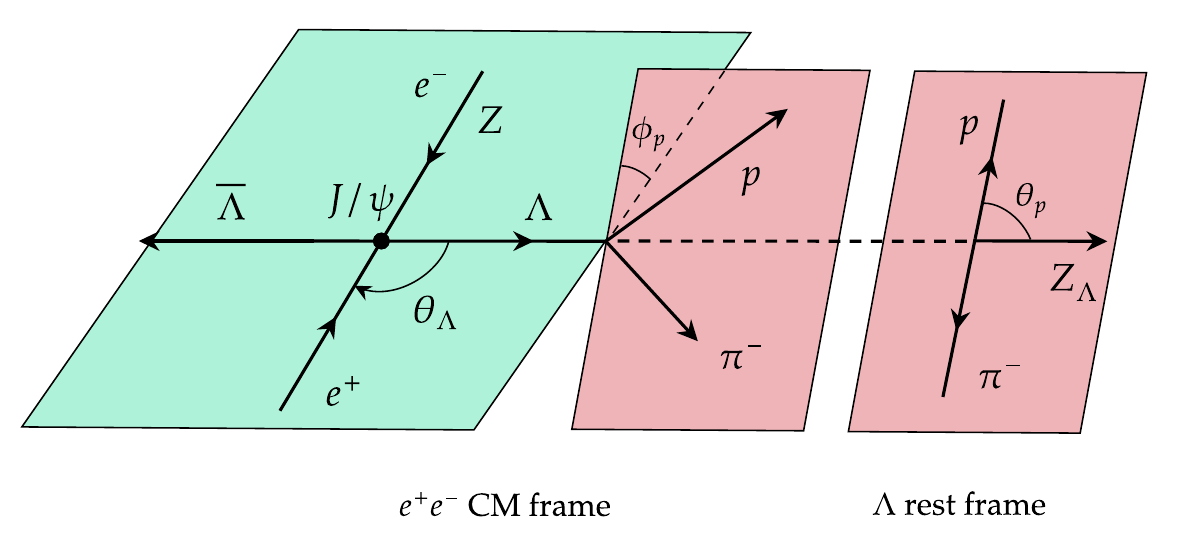} 
\caption{Definition of the coordinate system for the $e^+e^- \rightarrow J/\psi \rightarrow \Lambda(\rightarrow p\pi^-)\overline{\Lambda}(\rightarrow \overline{p}\pi^+)$.} 
\label{fig:coordinate}
\end{figure}

For the decay $J/\psi \rightarrow \Lambda(\rightarrow p\pi^-)\overline{\Lambda}(\rightarrow \overline{p}\pi^+)$, the entire process can be completely specified by the five-dimensional kinematic vector $\xi = (\theta_\Lambda, \theta_p, \phi_p, \theta_{\bar{p}}, \phi_{\bar{p}})$, where the corresponding helicity frames and coordinate systems are defined in Figure~\ref{fig:coordinate}.

To define these angles, a right-handed Cartesian coordinate system is established. In the rest frame of the $\Lambda$ hyperon, the $Z_\Lambda$ axis is aligned along the momentum direction of the $\Lambda$ in the $e^+e^-$ CM frame. The other two unit vectors are defined via cross products as $Y_\Lambda = Z \times Z_\Lambda$ and $X_\Lambda = Y_\Lambda \times Z_\Lambda$, in which $Z$ denotes the momentum direction of the electron beam in the $e^+e^-$ center-of-mass frame. For the $\overline{\Lambda}$ hyperon, the corresponding coordinate system is constructed symmetrically as $(-X_\Lambda, Y_\Lambda, -Z_\Lambda)$. Within these reference systems, $\theta_\Lambda$ denotes the production angle of the $\Lambda$ hyperon relative to the incoming electron beam direction in the $e^+e^-$ center-of-mass frame, and $(\theta_p, \phi_p)$ or $(\theta_{\bar{p}}, \phi_{\bar{p}})$ represent the polar and azimuthal angles of the motion directions of daughter proton $p$($\overline{p}$) in the $\Lambda$($\overline{\Lambda}$) rest frame.

The general spin density matrix for a baryon pair $B\bar{B}$ produced in $e^+e^-$ annihilation via an intermediate vector meson like $J/\psi$ can be formulated as:
\begin{equation}
R(\lambda_1, \lambda_2; \lambda'_1, \lambda'_2) \propto \sum_{m, m'} \rho_{m, m'} d^{j=1}_{m, \lambda_1 - \lambda_2}(\theta) d^{j=1}_{m', \lambda'_1 - \lambda'_2}(\theta) \mathcal{M}_{\lambda_1, \lambda_2} \mathcal{M}^*_{\lambda'_1, \lambda'_2} \delta_{m, m'}
\label{eq:R_amplitude}
\end{equation}
where the indices $m^{(\prime)}$ and $\lambda_{1,2}^{(\prime)}$ represent the helicities of the vector meson and $B(\bar{B})$ baryons, respectively. For $J/\psi$, $m^{(\prime)} = +, -, 0$. Since the $\rho_{00}$ component is heavily suppressed by $m_e^2/M_{J/\psi}^2$, the spin density matrix $\rho_{m,m'}$ for $J/\psi$ is simplified to a $2 \times 2$ matrix. The dynamics of $J/\psi \to B\bar{B}$ can be described with the helicity amplitude $\mathcal{M}_{\lambda_1^{(\prime)},\lambda_2^{(\prime)}}$. And the $d^{j}_{m^{(\prime)},\lambda_1^{(\prime)}-\lambda_2^{(\prime)}}(\theta)$ is the Wigner rotation function. The polar angle $\theta$ corresponds to the $\theta_\Lambda$ as we mention before.

During the production process of $J/\psi$ via $e^+e^-$ annihilation, the interference between the contributions from virtual $\gamma$ and $Z$-boson exchanges leads to a small longitudinal polarization of the $J/\psi$ meson, which can be represented as the relative difference of the diagonal elements of $\rho_{m,m'}$:
\begin{equation}
P_L = \frac{\rho_{++} - \rho_{--}}{\rho_{++} + \rho_{--}}.
\label{eq:PL_definition}
\end{equation}
And the effects of this polarization will be encoded into the spin density matrix of $B\bar{B}$ pair.

For the decay process of $J/\psi$ to $\Lambda\overline{\Lambda}$, it involves the contributions from the strong interaction, quantum electrodynamics (QED), and weak forces within the SM. To investigate the P and \textit{CP} violation effects in this decay, the Lorentz-invariant helicity amplitude is written as the sum of four independent form factors fixed at $q^2 = M^2_{J/\psi}$~\cite{He:2023plb}: 
\begin{equation}
\mathcal{M}_{\lambda_1, \lambda_2} = \epsilon_\mu(\lambda_1 - \lambda_2) \overline{u}(\lambda_1, p_1) \left( F_V \gamma^\mu + \frac{i}{2m_\Lambda} \sigma^{\mu\nu} q_\nu H_\sigma + \gamma^\mu \gamma^5 F_A + \sigma^{\mu\nu} \gamma^5 q_\nu H_T \right) v(\lambda_2, p_2),
\label{eq:amplitude}
\end{equation}
where $m_\Lambda$ is the mass of $\Lambda$, and $p_1, p_2$ are the four-momenta of $\Lambda$ and $\overline{\Lambda}$ in $J/\psi$ rest frame. 

In Eq.~(\ref{eq:amplitude}), $F_V$ and $H_\sigma$ represent the Dirac and Pauli form factors, which govern the \textit{CP}-conserving strong and electromagnetic production mechanisms. They can mathematically be related to the redefined magnetic and electric form factors $G_1$ and $G_2$~\cite{BESIII:2019natphys}:
\begin{equation}
F_V = G_1 - 4m^2 \frac{(G_1 - G_2)}{(p_1 - p_2)^2}, \quad H_{\sigma} = 4m^2 \frac{(G_1 - G_2)}{(p_1 - p_2)^2}.
\label{eq:emformfactor}
\end{equation}
Then two experimental observables, angular decay asymmetry parameter $\alpha_{J/\psi}$ and the relative strong phase $\Delta\Phi$, can be introduced and used to describe the polarization of $\Lambda$ via~\cite{BESIII:2022prl}:
\begin{equation}
\alpha_{J/\psi} = \frac{M^2_{J/\psi} |G_1|^2 - 4m^2_\Lambda |G_2|^2}{M^2_{J/\psi} |G_1|^2 + 4m^2_\Lambda |G_2|^2} \quad \text{and} \quad \frac{G_1}{G_2} = \left| \frac{G_1}{G_2} \right| e^{-i\Delta\Phi}.
\label{eq:observables}
\end{equation}

The remaining two form factors, $F_A$ and $H_T$, encapsulate the symmetry-violating mechanisms within the amplitude. Specifically, the $F_A$ represents the parity-violating contribution arising from the $Z$-boson exchange in the $J/\psi \rightarrow \Lambda\overline{\Lambda}$ decay. Meanwhile, the tensor form factor $H_T$, acting as the electric dipole form factor, characterizes the \textit{CP} violating effects in the decay. According to Ref.~\cite{He:1993,He:1994}, the $\Lambda$ EDM is assumed to be the dominant source of this electric dipole form factor. And the relationship between $H_T$ and the EDM can be expressed as Eq.~(\ref{eq:ht_edm}), when neglecting the momentum transfer $q^2$ dependence of $H_T$~\cite{He:1993, He:1994, He:2023plb}:
\begin{equation}
H_T = \frac{2e}{3M_{J/\psi}^2} g_V d_\Lambda,
\label{eq:ht_edm}
\end{equation}
Here, $e$ denotes the magnitude of the electron charge and $g_V$ can be calculated from the $J/\psi \rightarrow e^+e^-$ decay width~\cite{He:1993}.

Based on the rotational symmetry, the joint differential angular distribution function for the complete decay chain is written as:
\begin{align}
\frac{d\sigma}{d\Omega} \propto \sum_{[\lambda], m, m'} & R(\lambda_1, \lambda_2; \lambda'_1, \lambda'_2) \nonumber \\
& \times D^{1/2*}_{\lambda_1, \lambda_p}(\phi_p, \theta_p) D^{1/2}_{\lambda'_1, \lambda'_p}(\phi_p, \theta_p) \mathcal{H}_{\lambda_p} \mathcal{H}^*_{\lambda'_p} \nonumber \\
& \times D^{1/2*}_{\lambda_2, \lambda_{\bar{p}}}(\phi_{\bar{p}}, \theta_{\bar{p}}) D^{1/2}_{\lambda'_2, \lambda'_{\bar{p}}}(\phi_{\bar{p}}, \theta_{\bar{p}}) \bar{\mathcal{H}}_{\lambda_{\bar{p}}} \bar{\mathcal{H}}^*_{\lambda'_{\bar{p}}},
\label{eq:joint_angular_dist}
\end{align}
where $[\lambda]$ is a set of all possible helicities of $\Lambda$, $\bar{\Lambda}$, $p$ and $\bar{p}$. The $D$ represents the Wigner rotation matrices, where $\xi = (\theta_\Lambda, \theta_p, \theta_{\overline{p}}, \phi_p, \phi_{\overline{p}})$ corresponds to a set of helicity angles mentioned before. The $\mathcal{H}$ is helicity amplitudes of the weak decay $\Lambda \rightarrow p\pi^-$, which contains the dynamics of the decay process, and can be used to define the parity violation parameter $
\alpha_\Lambda = \frac{|\mathcal{H}_{+1/2}|^2 - |\mathcal{H}_{-1/2}|^2}{|\mathcal{H}_{+1/2}|^2 + |\mathcal{H}_{-1/2}|^2}$.

\section{ANGULAR DISTRIBUTION FITTING AND SENSITIVITY}
\label{sec:fitresult}
\subsection{Selection Criteria}
\begin{figure}[htpb]
  \centering
  \begin{subfigure}[b]{0.48\linewidth}
    \raggedright \textbf{(a)} \\
    \centering
    \includegraphics[width=\textwidth]{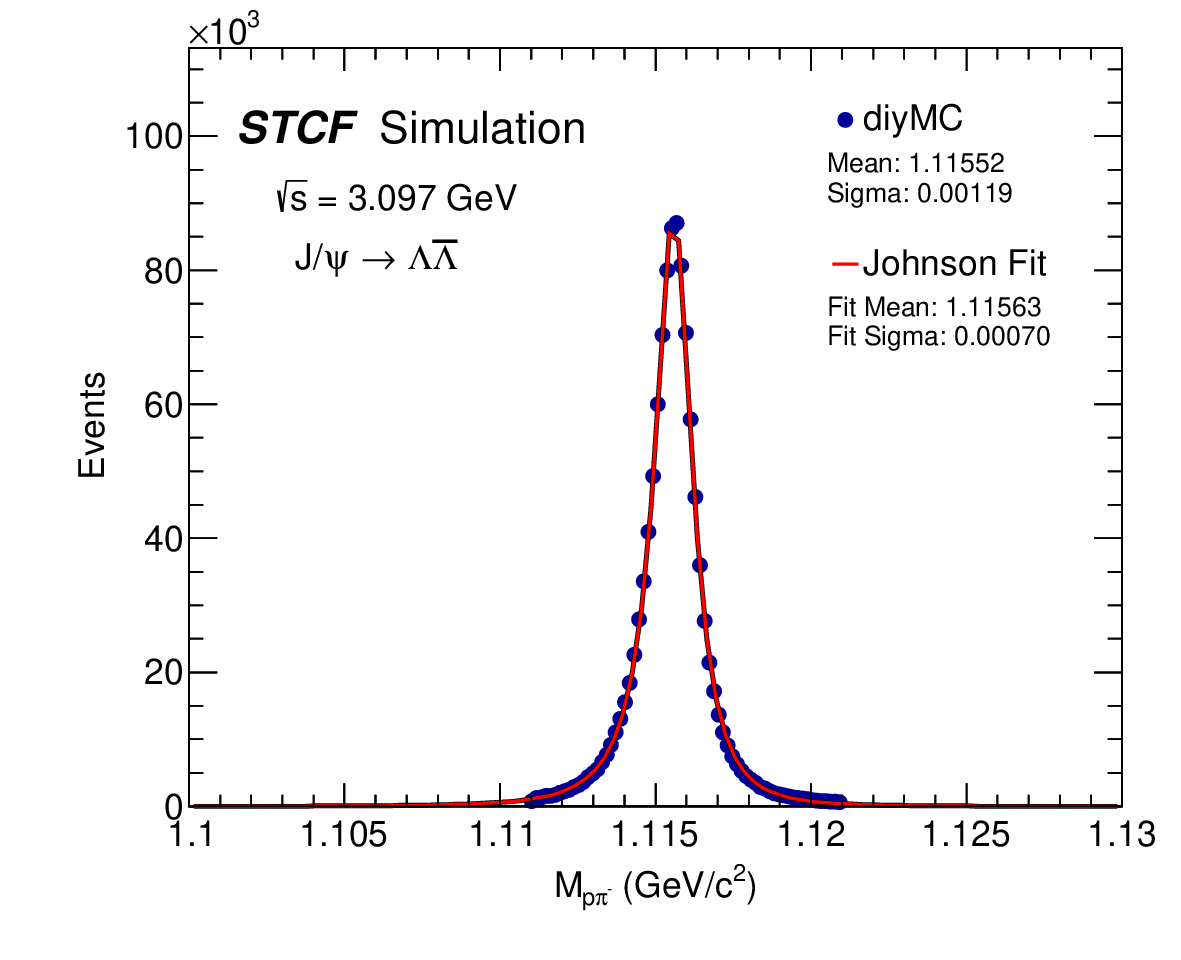}
  \end{subfigure}
  \hfill
  \begin{subfigure}[b]{0.48\linewidth}
    \raggedright \textbf{(b)} \\
    \centering
    \includegraphics[width=\textwidth]{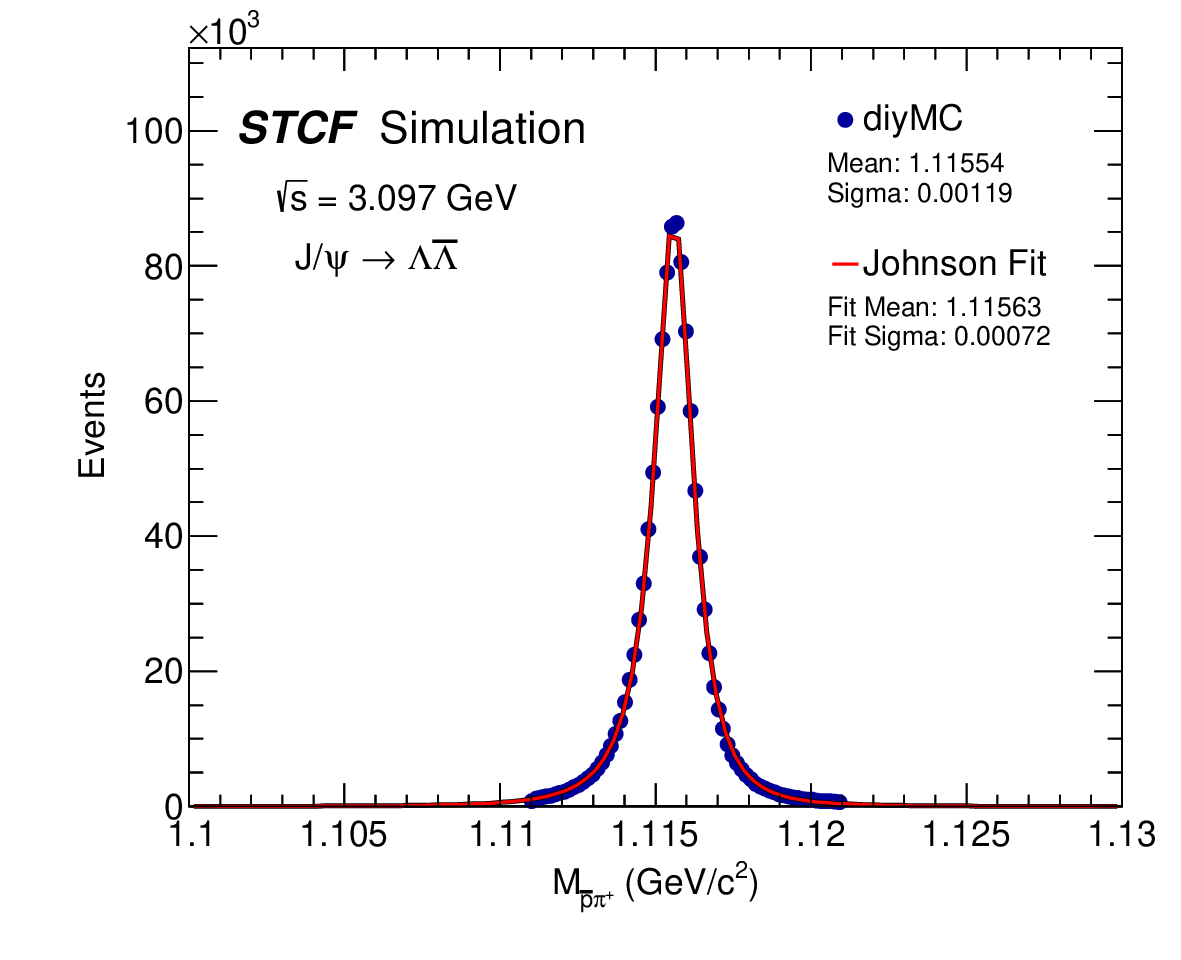}
  \end{subfigure}
  
  \caption{The invariant mass spectra for $\Lambda$ (a) and $\bar{\Lambda}$ (b) candidates after applying the selection criteria. The distribution is parameterized by a Johnson function.}
  \label{mass_of_lambda}
\end{figure}

The final event selection criteria are established as follows: Good charged tracks detected in the tracking system are required to be within a polar angle($\theta$) range of $|\cos\theta| < 0.94$. For the decay process $J/\psi \rightarrow \Lambda\overline{\Lambda}$ where $\Lambda$ and $\overline{\Lambda}$ decay into the charged final states $p\pi^-$ and $\overline{p}\pi^+$, the total number of good charged tracks in an event must be at least four. Protons and pions are separated clearly by a momentum threshold of $0.5$~GeV/$c$, above which particles are identified as protons, and below as pions. The $\Lambda(\overline{\Lambda})$ candidates are reconstructed by combining all $p\pi^-$($\bar{p}\pi^+$) candidates and then constraining the production position of $p\pi^-$($\bar{p}\pi^+$) to a common vertex through the vertex fit. Subsequently, a secondary vertex fit is applied to fit the decay length of $\Lambda(\overline{\Lambda})$, and the $p\pi^-$($\bar{p}\pi^+$) combination with the minimum $\chi^2_{\text{svtx}}$ is selected. Besides, the decay length of $\Lambda(\overline{\Lambda})$ must be positive. Finally, an energy-momentum conservation constraint is performed on the final selected $p\pi^-$($\bar{p}\pi^+$) hypothesis through the kinematic fit. The $\chi^2_{4\text{C}}$ of kinematic fit is required to be less than 100 and the invariant mass of $M_{p\pi}$ is required to be within $(1.111, 1.121)$~GeV/$c^2$. The invariant mass distributions of $\Lambda$ and $\bar{\Lambda}$ after applying all selection criteria are shown in Fig.~\ref{mass_of_lambda}. To better accommodate the distribution of the mass spectra, the signal is modeled using a four-parameter Johnson function~\cite{Johnson:1949zj}, offering enhanced flexibility over a conventional Gaussian distribution. After all these selection criteria, the final event selection efficiency is determined to be $37.55\%$.

\subsection{Angular Distribution Fitting}
To extract a set of parameters of interest, a maximum likelihood fit is performed based on the differential angular distribution shown in Equation~\ref{eq:joint_angular_dist}. And the likelihood function is constructed as:
\begin{equation}
\mathcal{L} = \prod_{i=1}^{N} \mathcal{P}(\xi^i; \vec{\lambda}) = \prod_{i=1}^{N} C W(\xi^i; \vec{\lambda}) \epsilon(\xi^i),
\label{eq:likelihood}
\end{equation}
Here, $\mathcal{P}(\xi^i; \vec{\lambda})$ represents the probability density function for event $i$, where $\xi^i$ is the kinematic variable of this event and $\vec{\lambda} = (\text{Re}(G_{2}), \allowbreak \text{Im}(G_{2}), \allowbreak P_{L}, \allowbreak \text{Re}(F_{A}), \allowbreak \text{Im}(F_{A}), \allowbreak \text{Re}(H_{T}), \allowbreak \text{Im}(H_{T}), \allowbreak \alpha_{\Lambda}, \allowbreak \alpha_{\bar{\Lambda}})$ represents the set of parameters to be fitted. The joint angular distribution formula $W(\xi^i; \vec{\lambda}) \propto d\sigma/d\Omega$ represents the weights assigned to each event. The detection efficiency is represented by $\epsilon(\xi^i)$, and N denotes the total number of events after the event selection. The normalization factor $C^{-1}$, denoted as $C^{-1} = \frac{1}{N_{\text{MC}}} \sum_{j=1}^{N^{\text{'}}} W(\xi^j; \vec{\lambda}) \epsilon(\xi^j)$, is calculated utilizing the generated $N^\text{'}$ PHSP MC sample, where $N_{MC}$ is the number of surviving events under the same selection criteria as the signal MC. To find the optimal parameter values, the parameter estimation is performed by minimizing the negative log-likelihood function, which serves as the final objective function:
\begin{equation}
\mathcal{S} = -\ln \mathcal{L},
\label{eq:objfuc}
\end{equation}
The minimization is implemented via the \textsc{RooFit}~\cite{Verkerke:2003econf} package with the MINUIT~\cite{Silin:1971cern} commands \textsc{MIGRAD} and \textsc{HESSE} in succession. During the fitting, the $G_1$ is fixed as a baseline scale, with $\text{Im}(G_1) = 0$ and $\text{Re}(G_1) = 1.61 \times 10^{-3}$, which is calculated from the measured decay rate of $J/\psi \to \Lambda\bar{\Lambda}$ and $\alpha_{J/\psi}$~\cite{Du:2024prd, PDG:2024}.

After the fitting, the sensitivities of the generated $3 \times 10^6$ signal MC can be obtained. Then based on this statistical precision, we estimate the statistical sensitivities that the future STCF experiment is expected to achieve. The STCF is expected to generate $3.4 \times 10^{12}$ $J/\psi$ events per year. Combined with the branching fractions $\mathcal{B}(J/\psi \rightarrow \Lambda\overline{\Lambda}) = 1.88 \times 10^{-3}$ and $\mathcal{B}(\Lambda \rightarrow p\pi) = 0.641$~\cite{PDG2026}, the expected number of $J/\psi \to \Lambda(\rightarrow p\pi^-)\overline{\Lambda}(\rightarrow \overline{p}\pi^+)$ events is about $2.63 \times 10^9$ per year. Consequently, the scale factor of statistical uncertainties can be calculated as $\sqrt{(3\times10^6)/(2.63\times10^9)} \approx 0.0338$. The projected statistical sensitivities are summarized in Table~\ref{tab:fit_result}, and the corresponding correlation coefficients are listed in Table~\ref{tab:coeff}. 

\begin{table}[htbp]
    \centering
    \caption{Fitting results for the angular distribution parameters. For STCF, only the statistical uncertainties are shown. 
    The parameters $\text{Re}(G_{2})$, $\text{Im}(G_{2})$, $P_{L}$, $\text{Re}(F_{A})$, $\text{Im}(F_{A})$, 
    $\text{Re}(H_{T})$, $\text{Im}(H_{T})$, $\alpha_{\Lambda}$ and $\alpha_{\bar{\Lambda}}$ are 
    directly related to the fitting and can be obtained directly from the fit, 
    while the parameters $\alpha_{J/\psi}$, $\Delta\Phi$, $A_{CP}$,  
    $\text{Re}(d_{\Lambda})$ and $\text{Im}(d_{\Lambda})$ are calculated by recombining the previously mentioned parameters.}
    \resizebox{0.75\textwidth}{!}{  
        \begin{tabular}{c|c|c}
            \hline
            \textbf{Parameters} & \textbf{\begin{tabular}[c]{@{}c@{}}BESIII Result~\cite{BESIII:2025edm}\\ ($1.0 \times 10^{10}\ J/\psi$)\end{tabular}} & \textbf{\begin{tabular}[c]{@{}c@{}}Projected Sensitivity at STCF\\ ($3.4 \times 10^{12}\ J/\psi$)\end{tabular}} \\ 
            \hline 
            $\alpha_{\Lambda}$       & $0.7524 \pm 0.0036 \pm 0.0008$ & $2.0 \times 10^{-4}$ \\ \hline 
            $\alpha_{\bar{\Lambda}}$ & $-0.7571 \pm 0.0036 \pm 0.0008$ & $2.0 \times 10^{-4}$ \\ \hline 
            $\text{Re}(G_{2})$       & $(9.71 \pm 0.06 \pm 0.24) \times 10^{-4}$ & $2.964 \times 10^{-7}$ \\ \hline 
            $\text{Im}(G_{2})$       & $(9.14 \pm 0.04 \pm 0.23) \times 10^{-4}$ & $2.227 \times 10^{-7}$ \\ \hline 
            $P_{L}$                  & $(-1.8 \pm 1.2 \pm 0.8) \times 10^{-3}$  & $6.7 \times 10^{-5}$ \\ \hline 
            $\text{Re}(F_{A})$       & $(-2.4 \pm 1.6 \pm 3.1) \times 10^{-6}$  & $8.7 \times 10^{-8}$ \\ \hline 
            $\text{Im}(F_{A})$       & $(-7.9 \pm 3.7 \pm 2.5) \times 10^{-6}$  & $2.1 \times 10^{-7}$ \\ \hline 
            $\text{Re}(H_{T})$       & $(-1.4 \pm 1.4 \pm 0.2) \times 10^{-6}$  & $8.1 \times 10^{-8}$ \\ \hline 
            $\text{Im}(H_{T})$       & $(1.3 \pm 1.2 \pm 0.4) \times 10^{-6}$   & $6.3 \times 10^{-8}$ \\ \hline 
            \hline 
            $\alpha_{J/\psi}$        & $0.4748 \pm 0.0022 \pm 0.0017$ & $1.202 \times 10^{-4}$ \\ \hline 
            $\Delta\Phi$              & $0.7552 \pm 0.0042 \pm 0.0013$ & $2.275 \times 10^{-4}$ \\ \hline 
            $A_{CP}$                 & $(-3.1 \pm 4.6 \pm 1.1) \times 10^{-3}$  & $2.6 \times 10^{-4}$ \\ \hline 
            $\text{Re}(d_{\Lambda})$ & $(-3.1 \pm 3.2 \pm 0.5) \times 10^{-19}$ & $1.8 \times 10^{-20}$ \\ \hline 
            $\text{Im}(d_{\Lambda})$ & $(2.9 \pm 2.6 \pm 0.6) \times 10^{-19}$  & $1.4 \times 10^{-20}$ \\ \hline 
            \hline
        \end{tabular}
    }
    \label{tab:fit_result}
\end{table}

\begin{table}[htbp] 
    \centering 
    \caption{Correlation matrix of the fitted parameters.} 
    \label{tab:coeff} 
    \resizebox{\textwidth}{!}{
        \begin{tabular}{c|ccccccccc} 
        \hline 
        \textrm{Paras.} & $Im(F_{A})$ & $Im(H_{T})$ & $Im(G_{2})$ & $P_{L}$ & $Re(F_{A})$ & $Re(H_{T})$ & $Re(G_{2})$ 
        & $\alpha_{\Lambda}$ & $\alpha_{\bar{\Lambda}}$ \\ 
        \hline 
        $Im(F_{A})$       & 1.000 & -0.002 & 0.018 & -0.002 & -0.085 & 0.000 & 0.060 & 0.009 & -0.008 \\  
        $Im(H_{T})$       &       & 1.000  & -0.004 & 0.007  & 0.001  & 0.355 & 0.007 & -0.026 & -0.027 \\
        $Im(G_{2})$       &       &        & 1.000  & -0.003 & -0.062 & -0.000 & -0.413 & -0.047 & 0.036 \\
        $P_{L}$           &       &        &        & 1.000  & -0.003 & 0.016 & 0.005 & -0.014 & -0.016 \\
        $Re(F_{A})$       &       &        &        &        & 1.000  & -0.005 & -0.126 & 0.023 & -0.001 \\
        $Re(H_{T})$       &       &        &        &        &        & 1.000  & 0.009 & -0.049 & -0.050 \\
        $Re(G_{2})$       &       &        &        &        &        &        & 1.000  & 0.088 & -0.082 \\
        $\alpha_{\Lambda}$         &       &        &        &        &        &        &        & 1.000  & 0.852 \\
        $\alpha_{\bar{\Lambda}}$   &       &        &        &        &        &        &        &        & 1.000 \\
        \hline 
        \end{tabular} 
    }
\end{table}

\begin{figure}[h!]
\centering
\includegraphics[width=0.5\textwidth]{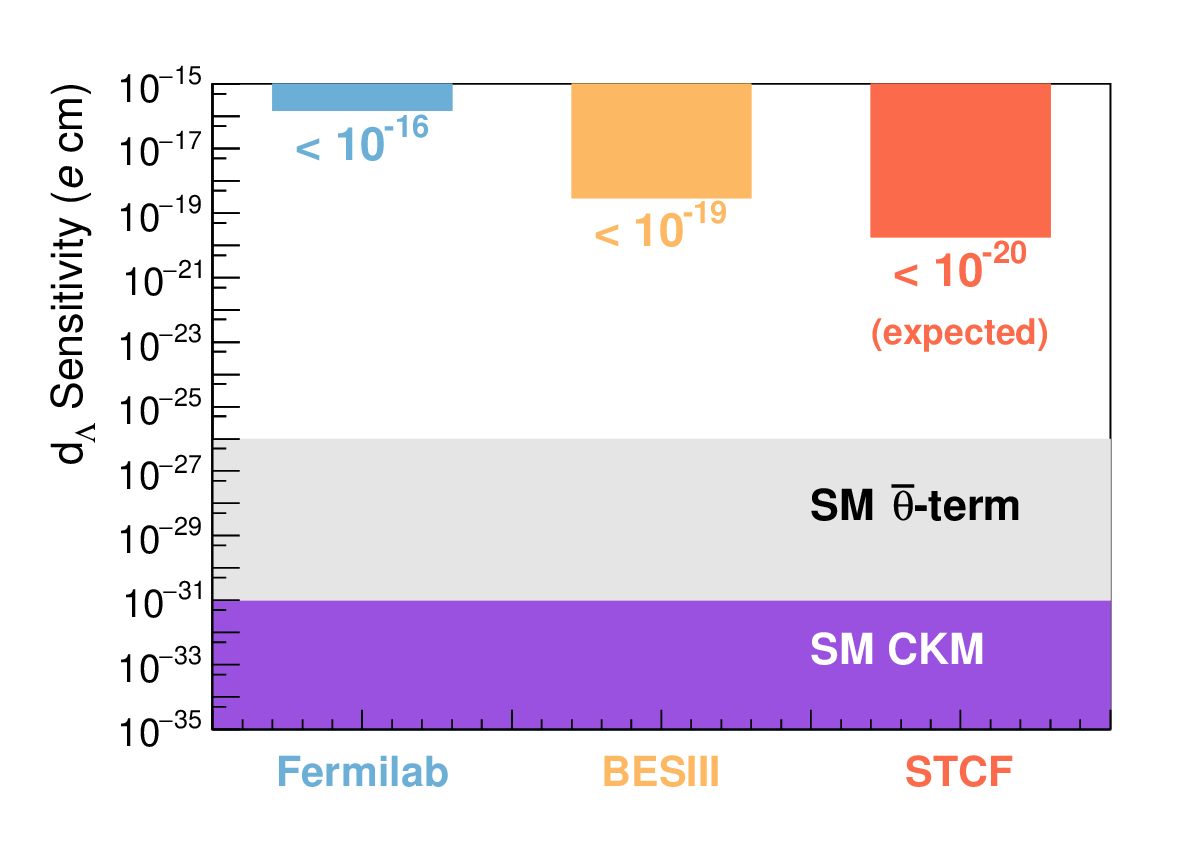} 
\caption{The upper limits of $d_\Lambda$ at different experiments.} 
\label{fig:edmcompare}
\end{figure}

We find that the statistical sensitivity for the direct \textit{CP} violation parameter $A_{CP}^{\Lambda}$ can reach $2.6 \times 10^{-4}$, while the sensitivity for the $\Lambda$ EDM can be probed at the $10^{-20}$~$e\text{ cm}$ level, with $\text{Re}(d_\Lambda) = 1.8 \times 10^{-20}$~$e\text{ cm}$ and $\text{Im}(d_\Lambda) = 1.4 \times 10^{-20}$~$e\text{ cm}$. These results represent a twofold improvement in precision compared to the current world-best results, highlighting the potential of the future STCF. And a comparison of the upper limits of $d_\Lambda$ between different experiments is shown in Figure~\ref{fig:edmcompare}, where the purple and gray part represent the theoretical predictions from the Standard Model CKM mechanism and the $\bar{\theta}$-term, respectively.

\begin{figure}[htbp]
  \centering
  % 第一张图：decay parameter (\alpha_-)
  \begin{subfigure}[b]{0.48\textwidth}
    \raggedright \textbf{(a)} \\ % 在图片外部左上角打印 (a)
    \centering
    \includegraphics[width=\textwidth]{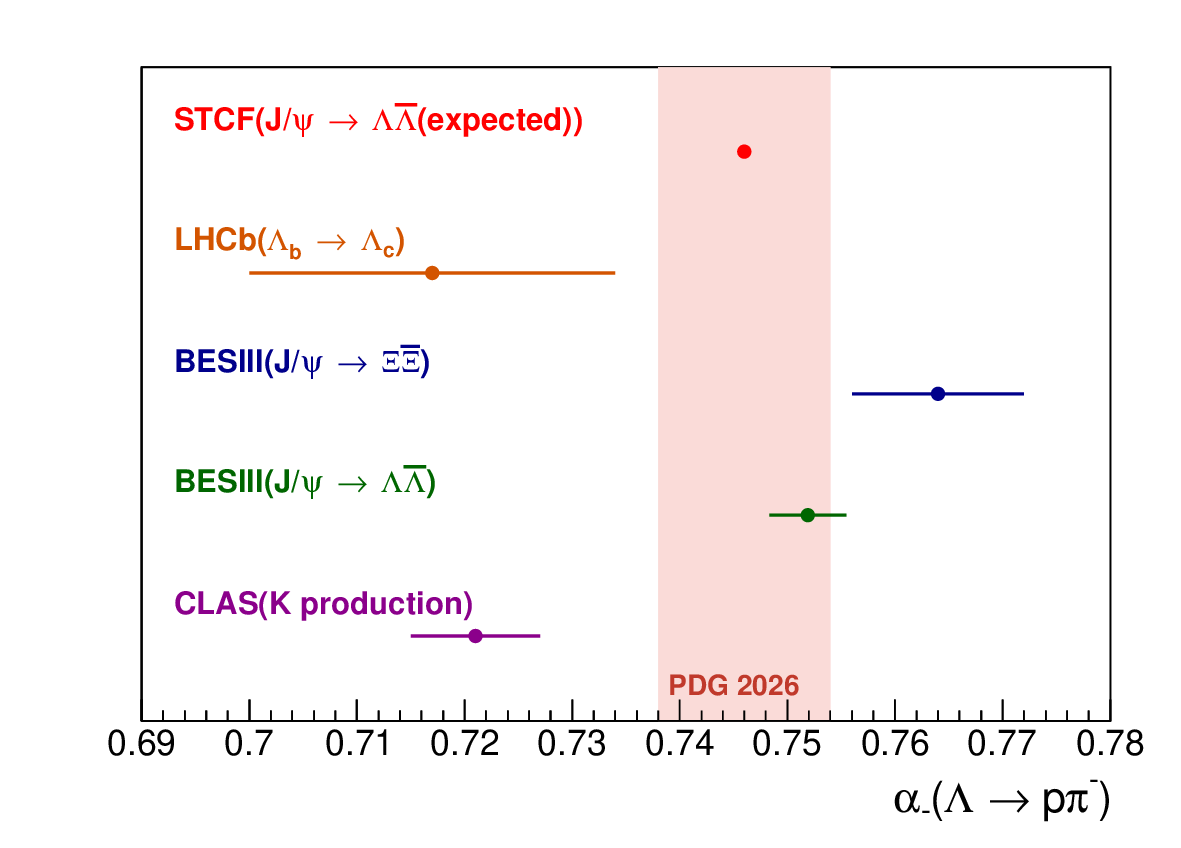}
    \label{fig:alpha_minus}
  \end{subfigure}
  \hfill % 弹性间距，将两张图推向两端并排
  % 第二张图：CP asymmetry (A_CP)
  \begin{subfigure}[b]{0.48\textwidth}
    \raggedright \textbf{(b)} \\ % 在图片外部左上角打印 (b)
    \centering
    \includegraphics[width=\textwidth]{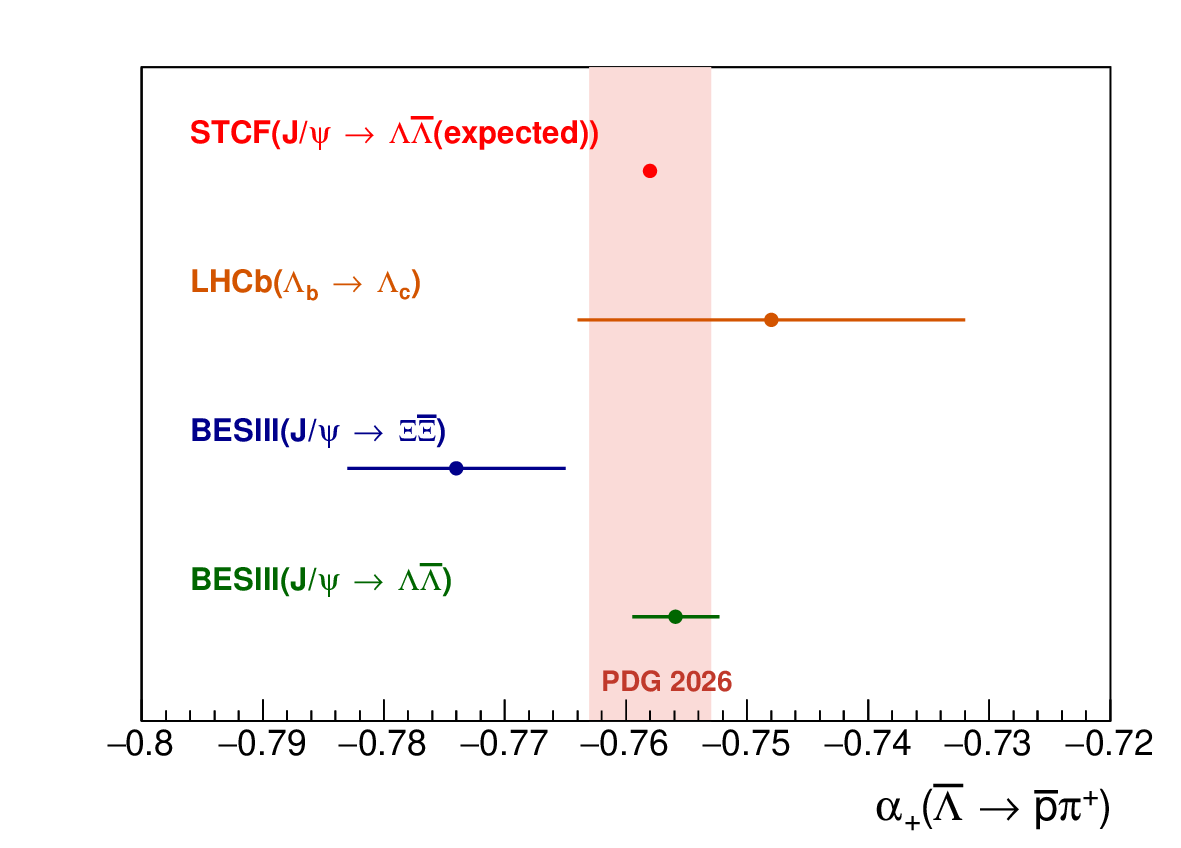}
    \label{fig:acp}
  \end{subfigure}
  
  \caption{Results of the $\Lambda$ decay parameter $\alpha_-$ (a) and $\bar{\Lambda}$ decay parameter $\alpha_+$ (b) at different experiments. The pink band represents the PDG 2026 value. And the data points represent the central values, while the error bars correspond to statistical uncertainties only. }
  \label{fig:decayparameter} % �� 注意：大图的 label 必须紧跟在大图的 \caption 后面！
\end{figure}

Results of the $\Lambda$ decay parameters and $A_{CP}^{\Lambda}$ from different experiments are shown in Figure~\ref{fig:decayparameter} and \ref{fig:acp}. Since the central value of the STCF cannot be determined at the current stage, it is set to match the current PDG value. Furthermore, due to the high precision achieved at STCF, the statistical uncertainty (of order $10^{-4}$) of STCF is barely visible in the figure.

\begin{figure}[h!]
  \centering
  \includegraphics[width=0.5\textwidth]{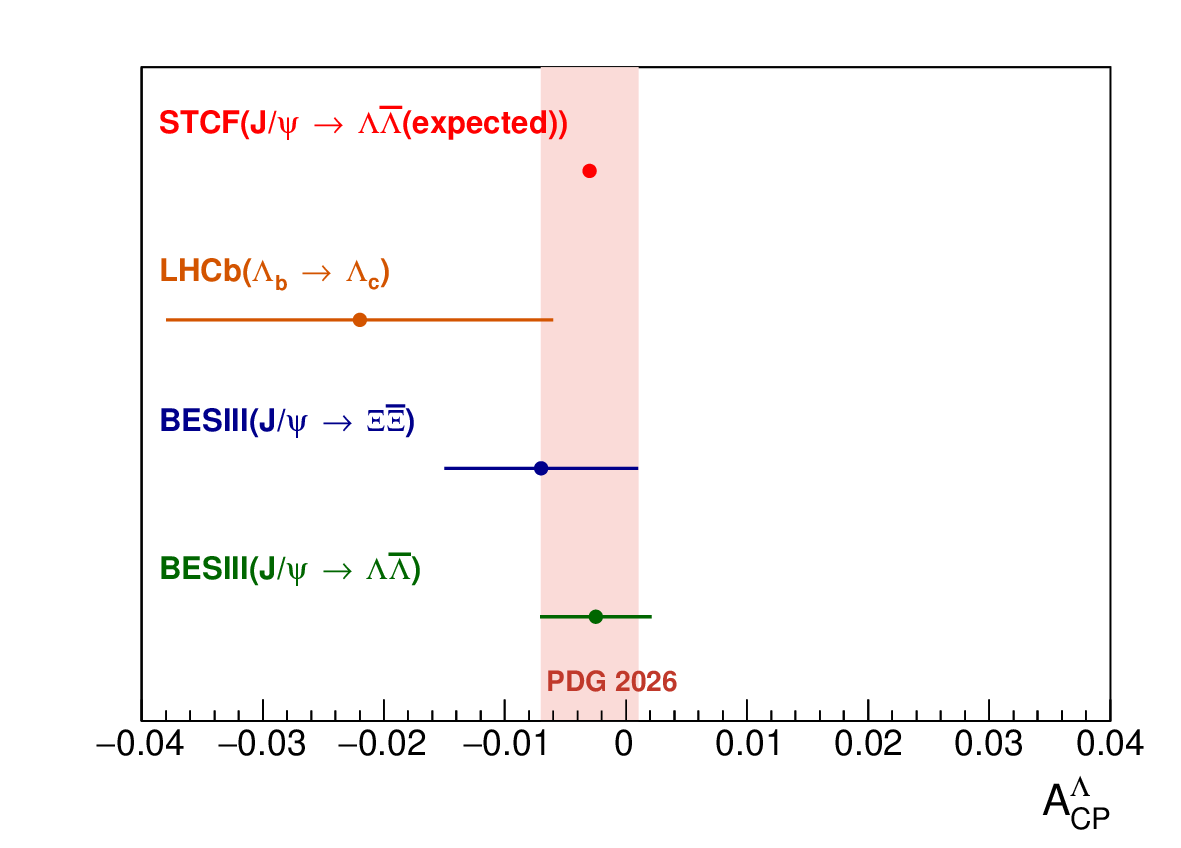}
  \caption{Results of $A_{CP}^{\Lambda}$ at different experiments. The data points represent the central values, while the error bars correspond to statistical uncertainties only. The pink band represents the PDG 2026 value.}
  \label{fig:acp}
\end{figure}

\begin{figure}[htpb]
    \centering
    \includegraphics[width=0.65\linewidth]{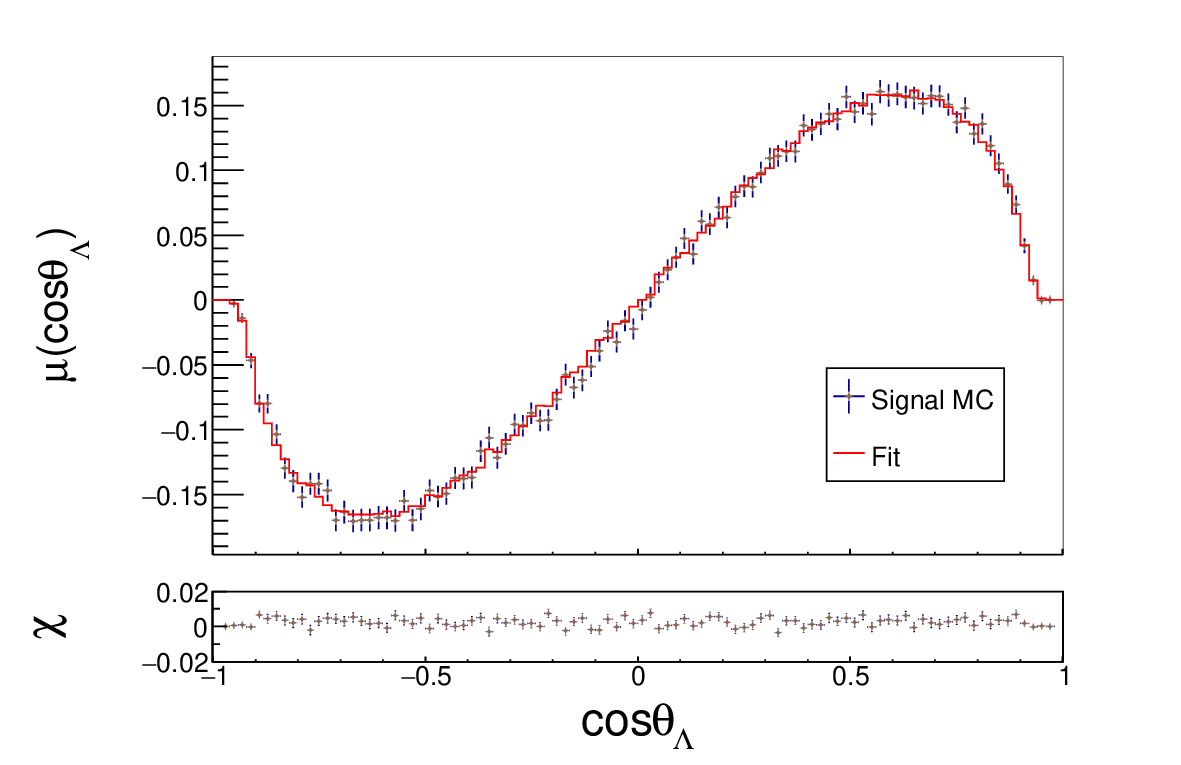}
    \caption{Distribution of the moment $\mu(\cos\theta_{\Lambda})$ as a function of $\cos\theta_{\Lambda}$. The points with error bars represent the signal MC sample, and the solid-line indicates the corresponding fit result. The lower panel displays the residual $\chi$ between the signal MC and the fit.}
    \label{fig:mu_distribution}
\end{figure}

To validate the reliability of the fit procedure, the consistency between the fit results and the generated signal MC sample is examined. Following Ref.~\cite{BESIII:2018cnd}, a moment $\mu(\cos\theta_{\Lambda})$ is constructed as
\begin{equation}
    \mu(\cos\theta_{\Lambda}) = \frac{m}{N} \sum^{N_k}_{i} \left( \sin\theta_p^i \sin\phi_p^i - \sin\theta_{\bar{p}}^i \sin\phi_{\bar{p}}^i \right),
\end{equation}
where $N$ is the total number of events after the final selection, $m=100$ is the number of bins in $\cos\theta_{\Lambda}$ used for calculating the moment, and $N_{k}$ denotes the number of events in the $k$-th $\cos\theta_{\Lambda}$ bin. The expected angular dependence of this moment, determined by the parameters obtained from the fit, is given by
\begin{equation}
    \mu(\cos \theta_{\Lambda}) = \frac{\alpha_{\Lambda}-\alpha_{\bar{\Lambda}}}{2} 
    \frac{1+\alpha_{J/\psi} \cos ^{2} \theta_{\Lambda}}{3+\alpha_{J/\psi}} P_{y}(\theta_{\Lambda}),
\end{equation}
where $P_{y}(\theta_{\Lambda})$ represents the transverse polarization of the $\Lambda$ hyperon, defined as $P_{y}(\theta_{\Lambda}) = \frac{\sqrt{1-\alpha_{J/\psi}^{2}}\cos\theta_{\Lambda}\sin\theta_{\Lambda}}{1+\alpha_{J/\psi}\cos^{2}\theta_{\Lambda}}\sin{\Delta\Phi}$. 
As illustrated in Fig.~\ref{fig:mu_distribution}, the moment distribution of the signal MC sample is in excellent agreement with the expected curve based on the fit results.

\section{PRELIMINARY SYSTEMATIC STUDY}
The major sources of systematic uncertainties in this analysis typically include background estimation, fitting procedure, detector resolution and data-MC inconsistencies, which originate from charged particle reconstruction, vertex reconstruction for $\Lambda(\bar{\Lambda})$ and energy-momentum conservation constraints. At the current stage, a direct evaluation of the inconsistency between the data and MC cannot be performed. In future studies, these inconsistencies will be comprehensively addressed using control samples to extract the efficiency differences between data and MC. Subsequently, the differences between the measurements with and without efficiency corrections are taken as the systematic uncertainty of this part. Furthermore, validation based on 500 pseudo-experiments in previous studies~\cite{BESIII:2025edm} has demonstrated that the fitting framework itself is entirely unbiased. Therefore, this study mainly investigates the systematic uncertainties originating from backgrounds estimation and detector resolution.

To evaluate the background systematic uncertainty, the main background is estimated based on $10^9$ $J/\psi$ inclusive MC sample at STCF. After applying the same selection criteria as for the signal MC, the purity of the signal is found to be 99.95\%. The dominant background processes primarily originate from $J/\psi \to \Lambda \bar{\Lambda} \gamma$, $J/\psi \to \pi^+ \pi^- p \bar{p}$, and $J/\psi \to \Lambda \bar{\Sigma}^0 + \mathrm{c.c.}$ (where the $\Sigma^0$ subsequently decays via $\Sigma^0 \to \Lambda \gamma$). To evaluate the systematic uncertainty, the normalized background samples (generated in Sec.~\ref{sec:stcfandmc}) are randomly mixed into the signal MC sample. Then the maximum likelihood fit is performed under two conditions: directly fitting the background-mixed sample, and fitting with the background subtracted. For the latter, the objective function is redefined as:
\begin{equation}
\mathcal{S} = -\ln\mathcal{L}_{\text{sig+bg}} + \ln\mathcal{L}_{\text{bg}}.
\label{eq:bg_subtraction}
\end{equation}
Finally, the resulting differences in the parameter mean values between these two fit scenarios are assigned as the systematic uncertainties of the background estimation.

Systematic uncertainties arising from the detector resolution are estimated utilizing toy-MC samples. The nominal ToyMC samples, which exclude resolution effects, were generated directly from Eq.~\ref{eq:joint_angular_dist}, where the amplitude parameters $\text{Re}(G_{2})$, $\text{Im}(G_{2})$, $P_{L}$, $\text{Re}(F_{A})$, $\text{Im}(F_{A})$, $\text{Re}(H_{T})$, $\text{Im}(H_{T})$, $\alpha_{\Lambda}$ and $\alpha_{\bar{\Lambda}}$ are fixed according to Refs.~\cite{BESIII:2025edm, BESIII:2022prl}. In total, 300 sets of nominal Toy MC samples were generated, with each containing $10^6$ events. Subsequently, the smeared toy-MC samples are generated by convolving the nominal distributions with the detector resolution, which are extracted from the discrepancies between the truth-level and reconstructed values of the five helicity angles, $\xi = (\theta_\Lambda, \theta_p, \theta_{\overline{p}}, \phi_p, \phi_{\overline{p}})$. To evaluate the systematic uncertainties, independent maximum-likelihood fits are performed on both the 300 nominal and 300 smeared toy-MC samples. 
The mean value of the difference distribution between the fitted values of the smeared and nominal sample pairs is assigned as the systematic uncertainty.

Conditioned on current STCF background projections and detector performance studies, systematic uncertainties associated with background and resolution for $A_{CP}^\Lambda$ and $\Lambda$ EDM are evaluated to be negligible.

\section{CONCLUSION}
Based on a projected dataset of $3.4 \times 10^{12}$ $J/\psi$ events at the future STCF experiment, we report the expected sensitivities for the \textit{CP} asymmetry parameters and the EDM in the quantum‑entangled $\Lambda\bar{\Lambda}$ pair, with the decay channel $J/\psi \rightarrow \Lambda(\rightarrow p\pi^-)\bar{\Lambda}(\rightarrow \bar{p}\pi^+)$. By using a novel full angular analysis method, the statistical uncertainty for the $CP$ asymmetry parameter $A_{CP}$ is estimated to be $2.6 \times 10^{-4}$, and the sensitivities for the EDM is estimated to be $\text{Re}(d_{\Lambda}) = 1.8 \times 10^{-20}$~$e\text{ cm}$ and $\text{Im}(d_{\Lambda}) = 1.4 \times 10^{-20}$~$e\text{ cm}$. Compared with the recent BESIII measurement, our projected statistical sensitivities show approximately one-order-of-magnitude improvement for both $A_{CP}$ and the $\Lambda$ EDM. We emphasize that these are purely statistical projections; final experimental uncertainties will also include systematic contributions. Nevertheless, these sensitivity projections at the STCF will place stringent constraints on new-physics parameter spaces in future, while also demonstrating the immense physics potential of the STCF in testing the SM and searching for new physics beyond it.

\section*{ACKNOWLEDGMENTS}
We thank the Hefei Comprehensive National Science Center for their strong support on the STCF key technology research project and the STCF software group for the development and maintenance of the simu-lation, reconstruction, and analysis software usedin this work. We are also grateful to the convenors of the STCF physics group, Cong Geng, Xiaolin Kang, Xiaorui Lyu, Xiaorong Zhou, and Yu Zhang, for their valuable suggestions and careful review of this work.

\end{document}